\documentclass[prd,aps,
nofootinbib,
floatfix,
superscriptaddress]{revtex4}
\usepackage{graphicx}
\usepackage{epsfig}
\usepackage{rotating}
\usepackage{amssymb}
\usepackage{subfigure}
\usepackage{dsfont}
\usepackage{psfrag}
\usepackage{amsmath,euscript,array,mathrsfs}
\usepackage{axodraw}

\newcommand{\SP}[1]{\begin{equation}\begin{split} #1
\end{split}\end{equation}}

\newcommand{\beq}{\begin{equation}}
\newcommand{\eeq}{\end{equation}}
\newcommand{\beqs}{\begin{eqnarray}}
\newcommand{\eeqs}{\end{eqnarray}}

\newcommand{\gsim}{\mathrel{\raisebox{-
.6ex}{$\stackrel{\textstyle>}{\sim}$}}}

\def\hbar{\hspace{0pt}\raisebox{1pt}{$-$} \hspace{-7pt} h}

\def\di{\mbox{d}}
\def\r{\rho}

\newcommand{\be}{\begin{equation}}
\newcommand{\ee}{\end{equation}}
\newcommand{\bea}{\begin{eqnarray}}
\newcommand{\eea}{\end{eqnarray}}

\def\lbldef#1#2{\expandafter\gdef\csname #1\endcsname {#2}}

\def\href#1#2{#2}

\newcommand{\ber}{\begin{eqnarray}}
\newcommand{\eer}{\end{eqnarray}}

\newcommand{\beqar}{\begin{eqnarray}}

\newcommand{\eeqar}{\end{eqnarray}}

\newcommand{\dsl}
  {\kern.06em\hbox{\raise.15ex\hbox{$/$}\kern-.56em\hbox{$\partial$}}}

\newcommand{\eeqarr}{\end{eqnarray}}
\newcommand{\ZZ}{{\rm \kern 0.275em Z \kern -0.92em Z}\;}

\def\CC{{\mathchoice
{\rm C\mkern-8mu\vrule height1.45ex depth-.05ex
width.05em\mkern9mu\kern-.05em}
{\rm C\mkern-8mu\vrule height1.45ex depth-.05ex
width.05em\mkern9mu\kern-.05em}
{\rm C\mkern-8mu\vrule height1ex depth-.07ex
width.035em\mkern9mu\kern-.035em}
{\rm C\mkern-8mu\vrule height.65ex depth-.1ex
width.025em\mkern8mu\kern-.025em}}}

\def\RR{{\rm I\kern-1.6pt {\rm R}}}

\def\ZZ{{\rm Z}\kern-3.8pt {\rm Z} \kern2pt}
\def\IB{\relax{\rm I\kern-.18em B}}
\def\ID{\relax{\rm I\kern-.18em D}}
\def\II{\relax{\rm I\kern-.18em I}}
\def\IP{\relax{\rm I\kern-.18em P}}

\newcommand{\bear}{\begin{eqnarray}}
\newcommand{\eear}{\end{eqnarray}}

\def\r{\rho}                                     

\def\6{\partial}

\def\bea{\begin{eqnarray}}
\def\eea{\end{eqnarray}}

\def\beqx{\begin{displaymath}}
\def\eeqx{\end{displaymath}}

\newcommand{\bmat}{\left(\begin{array}}
\newcommand{\emat}{\end{array}\right)}

\def\r{\rho}

\def\bo{{\raise-.3ex\hbox{\large$\Box$}}}               
\def\face{{\raise.2ex\hbox{$\displaystyle \bigodot$}\mskip-2.2mu \llap {$\ddot
        \smile$}}}                                   
\def\>{\rangle}                                      
\def\<{\langle}                                      

\def\leftrightarrowfill{$\mathsurround=0pt \mathord\leftarrow \mkern-6mu
        \cleaders\hbox{$\mkern-2mu \mathord- \mkern-2mu$}\hfill
        \mkern-6mu \mathord\rightarrow$}        
\def\dvec#1{\vbox{\ialign{##\crcr
        \leftrightarrowfill\crcr\noalign{\kern-1pt\nointerlineskip}
        $\hfil\displaystyle{#1}\hfil$\crcr}}}           

\def\-{\hphantom{-}}

\begin{document}
\title{On the glueball spectrum of walking backgrounds from wrapped-D5 gravity duals}

\author{Daniel Elander}
\affiliation{Department of Theoretical Physics, Tata Institute of Fundamental Research, \\
Homi Bhabha Road, Mumbai 400 005, India}

\author{ Maurizio Piai}

\affiliation{Department of Physics, College of Science, Swansea University,
Singleton Park, Swansea, Wales, UK.}

\date{\today}

\begin{abstract}

We compute the mass spectrum of glueball excitations of a special class of strongly-coupled field theories
via their type-IIB supergravity dual. We focus on two subclasses of backgrounds, which have different UV-asymptotics,
but both of which exhibit walking behavior, in the weak sense that the gauge coupling of the dual field theory exhibits a 
quasi-constant behavior at strong coupling over a  range of energies, before diverging in the deep IR.
We improve on earlier  calculations,
by  making use of the fully rigorous treatment of the 5-dimensional
consistent truncation, including the rigorous form of the boundary conditions. 
In both cases there is a parametrically light 
 scalar glueball. In the first case, this is a physical state,  while
  in the second case this result is unphysical, since the presence of higher-order operators in the dual  field theory
 makes the whole (physical) spectrum depend explicitly on a (unphysical) UV-cutoff scale.

\end{abstract}

\maketitle

\tableofcontents

\section{Introduction}

The study of strongly-coupled models of electroweak symmetry breaking (technicolor)
is notoriously difficult, because of the strongly-coupled dynamics itself~\cite{TC,reviewsTC}.
In particular, it has been the matter of a long debate in the literature whether
in a special class of such models, usually referred to as {\it walking}~\cite{WTC}, 
a parametrically light pseudo-dilaton exists in the spectrum~\cite{dilaton}, the phenomenology of which
would resemble that of an elementary Higgs particle~\cite{dilatonpheno,dilaton4,dilaton5D,dilatonnew}.
With the discovery at the LHC of a scalar particle with mass of approximately 126 GeV~\cite{ATLAS,CMS},
this fundamental field-theory question has become of utmost importance also from the
phenomenology perspective. If one could prove that no such light dilaton exists in any 
reasonable strongly-coupled theory, this theoretical fact, in combination with the experimental results,
might suggest  that electroweak symmetry breaking emerges in nature from a weakly-coupled sector.
Vice-versa, purely on the basis of the LHC data,  a number of phenomenological analyses~\cite{dilatonandpheno} 
show that  while the
 126 GeV scalar is compatible with the elementary Higgs of the Standard Model,
it could also be a composite dilaton emerging from strongly-coupled dynamics.

Gauge-gravity dualities~\cite{AdSCFT,reviewAdSCFT} offer an unprecedented opportunity to investigate 
the strong dynamics (a vast amount of lattice studies also appeared in recent years~\cite{lattice}), 
because they allow to rewrite untreatable strongly-coupled field theories in 
terms of  extra-dimensional theories of quantum gravity in a more accessible limit in which 
calculations are possible.
In particular, a completely algorithmic procedure exists for computing the spectrum of glueballs of 
strongly-coupled field theories, provided one knows not only the dual (10-dimensional) background, 
but also the 5-dimensional sigma-model action obtained by consistent truncation~\cite{BHM,EP}.
This algorithmic procedure makes crucial use of the diffeomorphism invariance of the 5-dimensional sigma-model action,
which allows to write the linearized equations for the fluctuations of the background in terms of physical (gauge-invariant) fields,
with well-defined boundary conditions.
Hence, it is nowadays possible to ask, within the framework of gauge-gravity duals, whether there are strongly-coupled, confining,
 field theories which admit a parametrically light state in the spectrum, to be identified as a dilaton.

The most challenging part of this program is to identify suitable candidate models.
It is known that within the effective field theory (EFT) framework (supplemented by naive dimensional analysis),
it is very difficult to write models in which parametrically light scalars exist, because of UV-sensitivity at the loop-level. 
In principle, there is no reason why such a scalar could not be light, but in practice this requires fine-tuning to
allow for cancellations between  large independent contributions (for a radically different and 
unconventional perspective on this point, see for instance~\cite{Sannino}). 
This is at the root of the hierarchy problem  in the context of models of electro-weak symmetry breaking, 
but the same types of EFT arguments  (and open problems) appear also in other contexts, 
for example in inflationary cosmology.
The mass of the light dilaton is not affected by such arguments in a fundamental (UV-complete) theory, 
in which the low-energy measurable quantities are not 
UV-sensitive. But then, one has to decide what kind of fundamental theory to study.
The physics lesson one learns from the EFT arguments is that the
 generic field theory is unlikely to produce a light dilaton, the existence of which would emerge only in  very
 special classes of dynamical models.
For this reason the number of candidate models (in the fully rigorous context of top-down approach to gauge-gravity dualities) 
is very limited, in spite of many studies~\cite{NPP,stringWTC,stringWTCbaryonic,stringWTC2}.
To the best of our knowledge, at present only one class of such models has been identified and studied in enough details,
and~\cite{ENP} reported on the presence of a parametrically light scalar state in the spectrum.
It is not known yet whether this state is a dilaton, nevertheless this result is of crucial importance: it can be thought of as
a proof of principle, stating that there exist non-trivial, confining strongly-coupled gauge theories in which 
a parametrically light scalar emerges from the dynamics.

A recent paper~\cite{ASW} could not find any light composite states.
The reader should be alerted of the fact  that the backgrounds considered in~\cite{ASW}
are those in~\cite{NPP} (we will call these backgrounds class 2 in the main body of the paper), 
which are   different from those in~\cite{ENP} (which we will refer to as class 1).
In this paper, we update the analysis of ~\cite{ENP}, refining the treatment of the boundary conditions and making it rigorous,
along the lines of ~\cite{EP}.
We confirm the main  results in~\cite{ENP}, showing that backgrounds of class 1 yield a parametrically light composite scalar in the spectrum.
We also perform the calculation in~\cite{ASW}, analyzing some special backgrounds of class 2, by using the rigorous boundary conditions, and without any of the approximations used in~\cite{ASW}. 
We show  not only that a light scalar 
is actually present in the spectrum also in class 2 (at least in some region of parameter space), but also that calculations in models of class 2 are difficult to interpret, since the whole spectrum of fluctuations is UV-cutoff dependent.
In the limit in which the UV-cutoff is removed, the theory has a continuous spectrum.
And on the other hand, keeping a finite cutoff means that solutions of this class admit an interpretation in
terms of a four-dimensional field theory that
is at best an effective field theory,
for which one expects radiative corrections to recreate the fine-tuning problems.
None of these problems arise in the first class of solutions, studied first in~\cite{ENP} and for which we refine the analysis in this paper.

The conclusion is that, after imposing the more rigorous (and easier to implement) boundary conditions of~\cite{EP}, we confirm the existence of  a light composite scalar,  
in important regions of parameter space, as originally stated in~\cite{ENP}. We also discuss several other open problems and possibilities, 
which we will address elsewhere.

\section{Generalities}

All the backgrounds of interest in this paper can be thought of as being generated by the 
strong-coupling limit of the background generated in type-IIB by a stack of $N_c$ $D5$-branes wrapping a special 2-cycle 
inside the base of the conifold, and hence are generalizations of the linear-dilaton solution in~\cite{MN}.
The five-dimensional description of the system is a subtruncation of the Papadopoulos-Tseytlin~\cite{PT}
truncation, itself a subtruncation of the most general consistent truncation of Type-IIB related to the conifold~\cite{consistentconifold}.

The coordinates are labelled by the Minkowski $x^{\mu}$, 
the radial direction $r$ (the only one that the ansatz will depend explicitly upon)
and five angles with ranges
$0\leq \theta,\tilde{\theta}<\pi\,,\,0\leq\phi,\tilde{\phi}<2\pi\,,\,0\leq\psi<4\pi$.  
The PT ansatz~\cite{PT} (setting $\alpha^{\prime}g_s=1$) is given by the following (Einstein frame) Type-IIB background functions:
\beqs
	ds^2&=&e^{2p-x} ds_5^2 + (e^{x+\tilde{g}} + a^2 e^{x-\tilde{g}}) (e_1^2 + e_2^2) + e^{x-\tilde{g}} \left( e_3^2 + e_4^2 - 2a (e_1 e_3 + e_2 e_4) \right) + e^{-6p-x} e_5^2, \\
	ds_5^2&=&dr^2 + e^{2A} dx_{1,3}^2, \\
	F_3&=& N \left[ e_5 \wedge \left(  e_4 \wedge e_3 + e_2 \wedge e_1 - b (e_4 \wedge e_1 - e_3 \wedge e_2) \right) + dr \wedge \left( \partial_r b ( e_4 \wedge e_2 + e_3 \wedge e_1) \right) \right], \\
	H_3&=&-h_2 e_5 \wedge (e_4 \wedge e_2 + e_3 \wedge e_1 ) + d r \wedge \Big[ \partial_r h_1 (e_4 \wedge e_3 + e_2 \wedge e_1 ) - \\ & & \partial_r h_2 (e_4 \wedge e_1 - e_3 \wedge e_2 ) + \partial_r \chi (-e_4 \wedge e_3 + e_2 \wedge e_1 ) \Big], \\
	F_5&=&\tilde F_5 + \star \tilde F_5, \ \ \tilde F_5 = -{\cal K} e_1 \wedge e_2 \wedge e_3 \wedge e_4 \wedge e_5,
\eeqs
where
\beqs
e_1&=&-\sin\theta \ d\phi\,,\\
e_2&=&d\theta\,,\\
e_3&=&\cos\psi\sin\tilde{\theta} \ d\tilde{\phi}\,-\,\sin\psi \ d\tilde{\theta}\,,\\
e_4&=&\sin\psi\sin\tilde{\theta} \ d\tilde{\phi}\,+\,\cos\psi \ d\tilde{\theta}\,,\\
e_5&=&d\psi+\cos\tilde{\theta} \ d\tilde{\phi}+\cos{\theta} \ d{\phi}\,,
\eeqs

together with the constraints
\beqs
	\mathcal K & = & M + 2N (h_1 + b h_2)\,,\\
	\partial \chi &=& \frac{(e^{2\tilde{g}}+2a^2+e^{-2\tilde{g}}a^4-e^{-2\tilde{g}})\partial 
	h_1+2a(1-e^{-2\tilde{g}}+a^2e^{-2\tilde{g}})\partial  h_2}{e^{2\tilde{g}}+(1-a^2)^2e^{-2\tilde{g}}+2a^2}\,.
\eeqs
All other fields of 10-dimensional Type IIB supergravity  vanish on the background.
The constants $M$ and $N$ control the fluxes of the $F_5$ and $F_3$  form, respectively.

We are interested only in solutions to the wrapped-$D5$ system, hence we set $M=0=h_1=h_2=\mathcal K =\chi$,
which leaves only the dilaton $\phi$, the metric and the $F_3$ form as non-trivial background functions.
In the following we pick $N_c = 4N = 1$. The background functions $(\tilde{g},p,x,\phi,a,b)$ depend only on the radial direction $r$.
The solutions we are interested in can be found by first making a change of variable $dr = 2 e^{-4p} d\rho$, and then putting~\cite{stringWTCbaryonic}
\SP{
	e^{2\tilde{g}} &= \frac{P^2-Q^2}{(P \coth(2\rho) - Q)^2}, \ \ 
	e^{-24p} = \frac{e^{4\phi_o}  (P^2 - Q^2) P'^3 \sinh(2\rho)}{131072}, \\
	e^{8x} &= \frac{e^{4\phi_o}  (P^2 - Q^2)^3 \sinh^2(2\rho)}{8192 P'}, \ \ 
	e^{4(\phi - \phi_o)} = \frac{\sinh^2 2\rho}{8(P^2 - Q^2)P'} \\
	a &= \frac{P}{\sinh(2\rho) (P \coth(2\rho) - Q)}, \ \ 
	b = \frac{2\rho}{\sinh(2\rho)} \\
	e^{6A} &= \frac{e^{4\phi_o}  (P^2-Q^2) \sinh^2(2\rho)}{256},
}
where
\SP{
	Q &= 2\rho \coth(2\rho) - 1\,.
}
Here, we set the end-of-space in the IR at $\r_0=0$, 
and in order to avoid a nasty singularity 
 in the IR we fine-tuned $Q_0=-N_c$ (the general form of the solution can be found in~\cite{HNP}, for example).

All the BPS equations of motion  are assured to be satisfied provided $P$ solves the ({\it master}) differential equation~\cite{HNP}
\SP{
	P'' + P' \left( \frac{P' - Q'}{P + Q} + \frac{P' + Q'}{P - Q} - 4 \coth(2\rho) \right) = 0.
}

The most general solution to the BPS equations is controlled by 5 integration constants.
 We fixed two of them already, and we will comment in due time about $\phi_o$,
 which appears in the 10-dimensional dilaton.
 The last two integration constants parameterize the general $P$.
 
 The solution $P=2N_c \r$ gives the background of~\cite{MN}. 
 Notice how this solution depends only on $N_c$, but there are no integration constants.
 There are two main classes of solutions, with different types of UV-asymptotic behavior.
\begin{itemize}
\item  Class 1: backgrounds where $P$ is approximately linear with $\r$ at asymptotically large $\r$. The solution in~\cite{MN} belongs to this class,
together with the class of walking backgrounds studied  in~\cite{ENP}. The field-theory interpretation of these solutions is explained 
for example in~\cite{stringWTCbaryonic}, and is connected with the infinite Higgsing of the dual quiver gauge group.
The dilaton is linear (in the UV). This class is characterized by one integration constant only,
namely the value of $P(0)=P_0$.
\item Class 2: backgrounds where $P\sim e^{4\r/3}$ diverges exponentially in the far-UV. The solutions discussed in~\cite{NPP} and studied by~\cite{ASW} belong to this class. As explained for example in~\cite{stringWTCbaryonic} (by drawing heavily on results from~\cite{quivers}), 
the dual field theory of these solution contains a dimension-8 operator, deforming the theory and making the UV badly behaved.
The dilaton is approximately constant in the far-UV. These solutions are characterized by two integration constants.
\end{itemize}

The Papadopoulos-Tseytlin~\cite{PT} ansatz can be rewritten by promoting all the functions appearing in the background
to eight five-dimensional scalar fields $\Phi^a$, with sigma-model action
defined by~\cite{BHM}
\beqs
{\cal S}&=&\int \di^5 x \sqrt{-g}\left(\frac{1}{4}R-\frac{1}{2}G_{ab}\partial_M\Phi^a\partial^M\Phi^b-V(\Phi) \right)\,,
\eeqs
with kinetic terms
\beqs
G_{ab}\partial_M\Phi^a\partial_N\Phi^b
&=&
\frac{1}{2}\partial_M \tilde{g} \partial_N \tilde{g} 
\,+\,\partial_M x \partial_N x
\,+\,6\partial_M p \partial_N p
\,+\,\frac{1}{4}\partial_M \Phi \partial_N \Phi \\
&&+\,\frac{1}{2}e^{-2\tilde{g}}\partial_M a \partial_N a
+\frac{1}{2}N^2e^{\Phi-2x}\partial_M b \partial_N b \nonumber\\
\nonumber &&
+\frac{e^{-\Phi-2x}}{e^{2\tilde{g}}+2a^2+e^{-2\tilde{g}}(1-a^2)^2}
\left[\frac{}{}(1+2e^{-2\tilde{g}}a^2)\partial_M h_1 \partial_N h_1\right.\\ &&\left.
+\frac{1}{2}(e^{2\tilde{g}}+2a^2+e^{-2\tilde{g}}(1+a^2)^2)\partial_M h_2 \partial_N h_2
+2a(e^{-2\tilde{g}}(a^2+1)+1)\partial_M h_1 \partial_N h_2\right]\,,\nonumber
\eeqs
and where the potential is
\beqs
V&=&-\frac{1}{2}e^{2p-2x}(e^{\tilde{g}}+(1+a^2)e^{-g})\\
&&\,+\,\frac{1}{8}e^{-4p-4x}(e^{2\tilde{g}}+(a^2-1)^2e^{-2\tilde{g}}+2a^2)\nonumber\\
&&\,+\,\frac{1}{4}a^2e^{-2\tilde{g}+8p}\nonumber\\
&&\,+\,\frac{1}{8}N^2e^{\Phi-2x+8p}\left[e^{2\tilde{g}}+e^{-2\tilde{g}}(a^2-2a b +1)^2 +2 (a-b)^2\right]\nonumber\\
&&\,+\,\frac{1}{4}e^{-\Phi-2x+8p}h_2^2\nonumber\\
&&\,+\,\frac{1}{8}e^{8p-4x}(M+2N(h_1+b h_2))^2\,.\nonumber
\eeqs
Again, in our case $h_1=h_2=M=0$ means that the last two lines of both the
kinetic terms and the potential drop.
The full five-dimensional PT system yields a vast space of possible background configurations,
in principle controlled by $16$ integration constants. This is a consistent truncation
of Type-IIB, in the sense that given any such background, one can construct the full 10-dimensional background
in a purely algebraic way, and what results is a solution of the classical equations of motion of Type-IIB.  
And the wrapped-$D5$ system (with six scalars) is itself a consistent truncation of PT.

\subsection{Fluctuations}

In order to compute the spectrum of scalar glueballs of the dual theory, one can study the fluctuations
of the background in the five-dimensional sigma-model. This will not yield the full spectrum,
but a subsector of it, which is going to contain the most important states at low energies.
In order to do so, one has to fluctuate all the scalars $\Phi^a$ remaining after truncating (six in our case),
but also the five-dimensional metric. 
We use the conventional ordering $\Phi^a = \{ \tilde{g},p,x,\phi,a,b \}$.
In doing so, one has to keep into account the fact that ${\cal S}$  is diffeomorphism invariant,
and hence many of the fluctuations are unphysical (pure gauge).
The problem is overcome by making use of appropriate combinations of 
the original fluctuations, constructed in such a way as to preserve gauge invariance in a manifest way.
These are called gauge invariant variables.
Following~\cite{BHM}, it turns out that the number of such propagating gauge-invariant fluctuations $\mathfrak{a}^a$
is the same as the number of independent scalars  in ${\cal S}$ (again, six for present purposes).

The fluctuations $\mathfrak a^a$ satisfy the following equation of motion in the bulk~\cite{EP}
\SP{\label{eq:fluceoms}
	\Big[ \mathcal D_r^2 + d A' \mathcal D_r + e^{-2A} q^2 \Big] \mathfrak{a}^a - \Big[ V^a_{\ |c} - \mathcal{R}^a_{\ bcd} \Phi'^b \Phi'^d + \frac{4 (\Phi'^a V_c + V^a \Phi'_c )}{(d-1) A'} + \frac{16 V \Phi'^a \Phi'_c}{(d-1)^2 A'^2} \Big] \mathfrak{a}^c = 0,
}
with boundary conditions
\SP{
\label{eq:BCs1}
	&\left[ \delta^a_{\ b} + e^{2A} q^{-2} \left( V^a - d A' \Phi'^a - \lambda^a_{\ |c} \Phi'^c \right) \frac{2 \Phi'_b}{(d-1) A'} \right] \mathcal D_r \mathfrak a^b \Big|_{r_i} = \\&  \left[ \lambda^a_{\ |b} + \frac{2 \Phi'^a \Phi'_b}{(d-1) A'} + e^{2A} q^{-2} \frac{2}{(d-1) A'} \left( V^a - d A' \Phi'^a - \lambda^a_{\ |c} \Phi'^c \right) \left( \frac{4 V \Phi'_b}{(d-1) A'} + V_b \right) \right] \mathfrak a^b \Big|_{r_i}.
}
The details about the derivation of these results can be found in~\cite{BHM,EP}.
Here, we only summarize the meaning of the notation: ${\cal D}_r$ is the covariant derivative with respect to the sigma-model metric $G_{ab}$,
$d=4$, $q^2$ relates to the Minkowski 4-momentum  with signature $\{+,-,-,-\}$, ${\cal R}^a_{\ bcd}$ is the Riemann tensor associated with the 
sigma-model metric $G_{ab}$,  which is also used to raise and lower the sigma-model indexes $a=1,\cdots,6$,
$V_a=\partial V/\partial \Phi^a$, primed quantities refer to derivatives in respect to $r$, and all the functions 
$A$, ${\cal R}$, $\Phi^a$, $V$ and their derivatives
are evaluated on the classical background. We refer the reader to ~\cite{BHM,EP} for more details about the formalism.

Notice how the boundary conditions depend on undetermined coefficients $\lambda^a_{\,\,\,\,|b}$ the meaning of which
is explained at length in~\cite{EP}.
In the limit of $\lambda^a_{\ |b} \big|_{r_i} \rightarrow \pm \infty$, which we adopt from now on, 
we obtain the boundary conditions
\SP{
\label{eq:BCs2}
	-e^{2A} q^{-2} \frac{2 \Phi'^a}{(d-1) A'} \left[ \Phi'_b \mathcal D_r  - \frac{4 V \Phi'_b}{(d-1) A'} - V_b \right] \mathfrak a^b \Big|_{r_i} = \mathfrak a^a \Big|_{r_i}.
}
The spectrum can be computed by solving the bulk equations and imposing the UV and IR boundary conditions
at two arbitrary end-of-space points in the radial direction, that we denote $r_1\ll r_2$. These serve as regulators.
The calculation has to be repeated for larger values of $r_2$ and smaller values of $r_1$ approaching the 
physical end of space. If the background is sufficiently well-behaved, this procedure converges,
and the final results do not depend on the regulators.

After changing the radial coordinate from $r$ to $\rho$, we can write Eq.~\eqref{eq:fluceoms} as (for $k = - A - 4p + \log 2$)
\SP{
\label{eq:flucdiffrho}
	\Big[ \delta^a_{\ b} \partial_\rho^2 + S^a_{\ b} \partial_\rho + T^a_{\ b} + \delta^a_{\ b} e^{2k} q^2 \Big] \mathfrak{a}^b = 0,
}
where
\SP{
	S^a_{\ b} =& 2 \mathcal{G}^a_{\ bc} \partial_\rho \Phi^c + 4 \left(\partial_\rho p + \partial_\rho A \right) \delta^a_{\ b}, \\
	T^a_{\ b} =& \partial_b \mathcal{G}^a_{\ cd} \partial_\rho \Phi^c \partial_\rho \Phi^d - 4 e^{-8p} \Bigg[ \left( \frac{4 (V^a \partial_\rho \Phi^c + \partial_\rho \Phi^a V^c)}{3 \partial_\rho A} + \frac{16 V \partial_\rho \Phi^a \partial_\rho \Phi^c}{9 (\partial_\rho A)^2} \right) G_{cb} + \partial_b V^a \Bigg].
}
We will see that this further rewriting is going to facilitate the analysis.

\section{Walking backgrounds with linear-dilaton asymptotics}

We start from the walking backgrounds of class 1. The spectrum has already been studied in~\cite{ENP}.
However, while making use of the gauge-invariant bulk equations, in that paper
the boundary conditions have been chosen on the basis of regularity arguments,
because the general form of the boundary conditions was not known at the time.
We are now in the position to redo this calculation, imposing the rigorous boundary conditions.

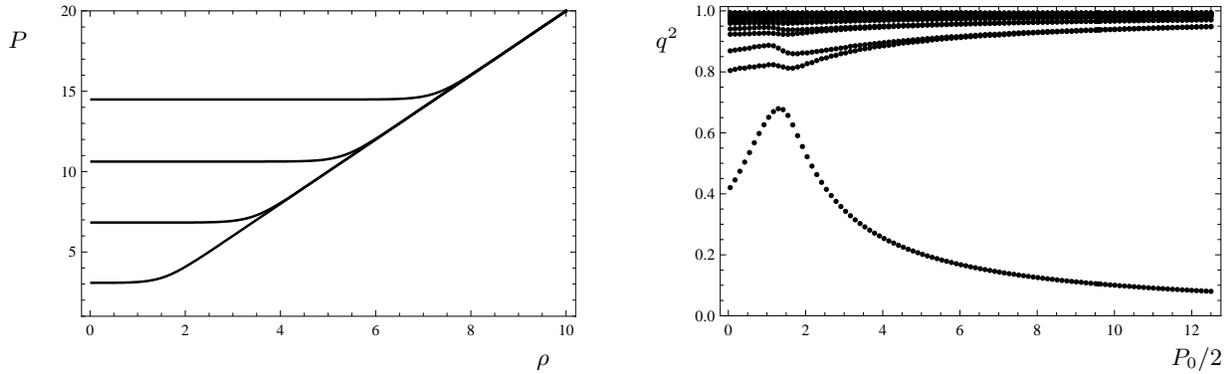
\begin{figure}[t]
\begin{center}
\begin{picture}(500,140)
\put(30,10){\includegraphics[height=4.5cm]{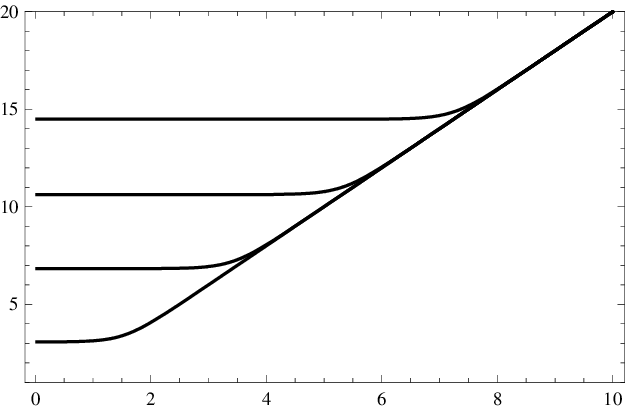}}
\put(270,10){\includegraphics[height=4.5cm]{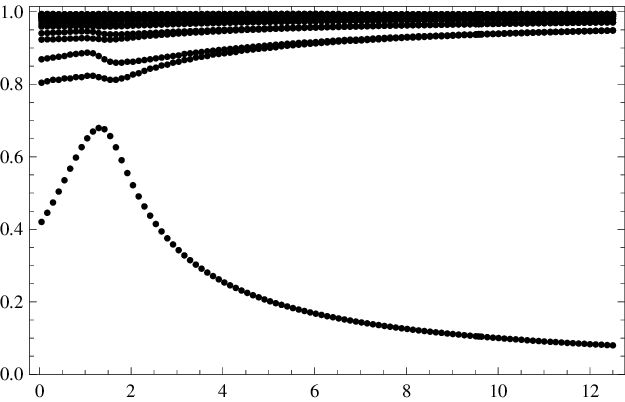}}
\put(10,120){$P$}
\put(255,120){$q^2$}
\put(210,0){$\rho$}
\put(450,0){$P_0/2$}
\end{picture}
\caption{The left panel shows $P$ for a few different backgrounds. The right panel shows the spectrum $q^2$ computed for different values of $P_0/2 \simeq \rho_*$. As $P_0 \rightarrow 0$, the spectrum agrees with the one for the solution in~\cite{MN}, as computed in~\cite{BHM}. 
For large $P_0$, there is a light scalar. These numerical results have been obtained with $\r_2=300$, and we checked that 
different (but large) values of this UV cutoff do not appreciably change the results.}
\label{Fig:spectrum1}
\end{center}
\end{figure}

\begin{figure}[t]
\begin{center}
\begin{picture}(500,140)
\put(30,10){\includegraphics[height=4.5cm]{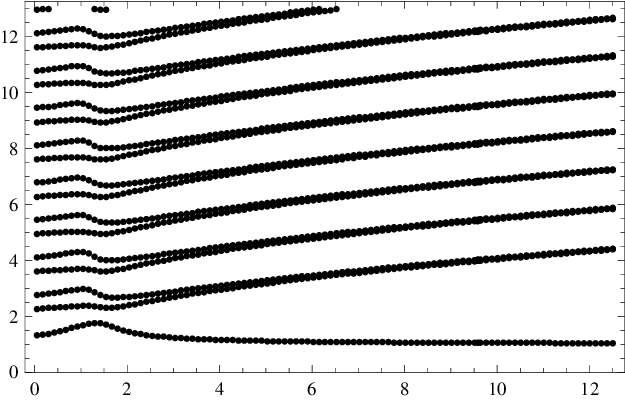}}
\put(270,10){\includegraphics[height=4.8cm]{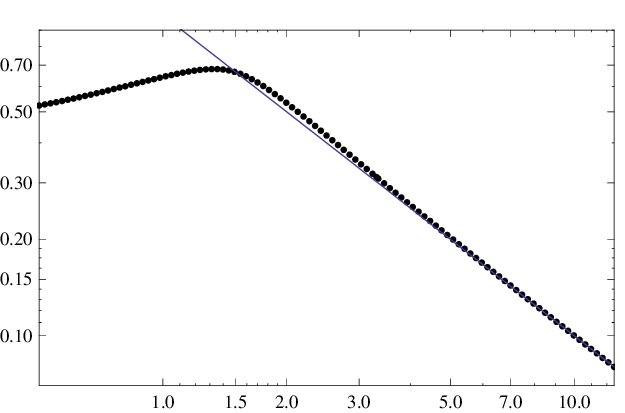}}
\put(0,120){$\frac{1}{\sqrt{1-q^2}}$}
\put(255,120){$q^2$}
\put(210,0){$P_0/2$}
\put(450,0){$P_0/2$}
\end{picture} 
\caption{The left panel shows $1 / \sqrt{1-q^2}$ versus $P_0/2 \simeq \rho_*$. The right panel shows a log-log plot of the mass squared of the lightest scalar versus $P_0/2 \simeq \rho_*$, with the blue line corresponding to $q^2 = 2/P_0$.}
\label{Fig:spectrum2}
\end{center}
\end{figure}
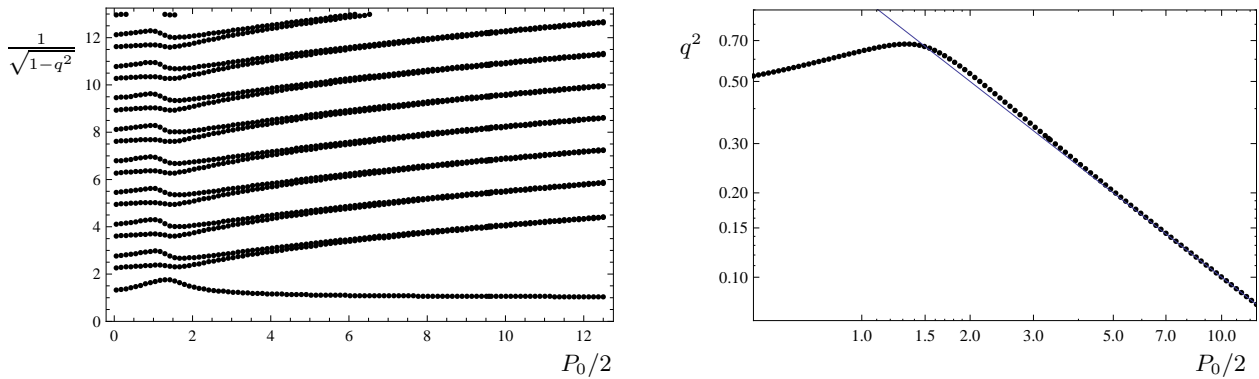

We proceed as follows. First of all, we choose a large class of solutions $P$ of the master equation 
which have linear UV-asymptotics (a few examples are shown in Fig.~\ref{Fig:spectrum1}).
All of them differ only by the value of $P(0)=P_0$, or, equivalently, by the value  $\r_{\ast}$ of the coordinate
at which the linear behavior appears, taking over from the approximately constant behavior of $P$ in the deep IR.
Then we introduce two cutoffs $\r_1$ and $\r_2$, and fix $\phi_o=0$. We compute the spectrum 
for all the backgrounds. We repeat this for various choices of $\r_2$ and $\r_1$, until the results show no dependence on the cutoffs.
The result is shown in Fig.~\ref{Fig:spectrum1}.
For completeness, we also show some more details about the spectrum in Fig.~\ref{Fig:spectrum2}.

Let us discuss the results.
First of all, the spectrum is discrete only up to $q=1$. For larger values of $q$ the system of linear equations breaks down:
while there are discrete solutions for finite $\r_2$, taking $\r_2\rightarrow \infty$ yields a continuous spectrum.
In order to understand where this comes from, let us expand in the far-UV:
\SP{
	P = 2\rho + \mathcal O\left( e^{-4\rho}\right).
}
Neglecting exponentially suppressed terms, this yields
\beq
	S^a_{\ b} = \left(
\begin{array}{cccccc}
 2 & 0 & 0 & 0 & 0 & 0 \\
 0 & 2 & 0 & 0 & 0 & 0 \\
 0 & 0 & 2 & 0 & 0 & 0 \\
 0 & 0 & 0 & 2 & 0 & 0 \\
 0 & 0 & 0 & 0 & 2 & 0 \\
 0 & 0 & 0 & 0 & 0 & 2
\end{array}
\right) + \mathcal O\left(\rho^{-1}\right), \ \ 
	T^a_{\ b} = \left(
\begin{array}{cccccc}
 -4 & 0 & 4 & -2 & 0 & 0 \\
 0 & -6 & -1 & -\frac{1}{2} & 0 & 0 \\
 2 & -6 & -3 & \frac{1}{2} & 0 & 0 \\
 -4 & -12 & 2 & -3 & 0 & 0 \\
 0 & 0 & 0 & 0 & -4 & 4 \\
 0 & 0 & 0 & 0 & 4 & -4
\end{array}
\right) + \mathcal O\left(\rho^{-1}\right),
\eeq
So that all the non-zero entries of these matrices are ${\cal O}(1)$ in the UV. But we also find that
\SP{
	e^{2k} = 1 + \mathcal O\left(e^{-4\rho}\right).
}
After going to a diagonal basis, one finds that the fluctuations behave (up to powers of $\rho$) as $\sim e^{\left( -1 \pm \sqrt{C-q^2} \right) \rho}$ 
with $C=1$ for half of them and $C=9$ for the other half.
This is the origin of the problem: we would like to interpret the (real) solutions of the bulk equations in terms of 
operators of the dual theory and their sources, but this cannot be done for large-$q^2$ because there is no
sense in which half of the asymptotic solutions are dominant and half subdominant, rather all the solutions become oscillatory in the far UV at large-$q$.
This is a well known behavior that appears in many quantum mechanics systems: provided one keeps the UV cutoff as physical, 
there is a discrete spectrum (similar to what happens for a potential well of finite size in quantum mechanics),
 but if the cutoff is removed, the spectrum becomes continuous.
The result is similar, qualitatively, to the hydrogen atom in quantum mechanics: a discrete spectrum at low energies, with the eigenvalues becoming closer and closer to each other until a critical value beyond which the spectrum is continuous.
We adopt the same interpretation: above $q=1$ the states are unbounded. The existence of a theory in which 
confinement yields a spectrum of bound states with an upper bound on the mass is at least unconventional.
But we do not need to worry about this: the low-energy behavior is consistent with
expectations.

Looking at the low-$q^2$ spectrum of excitations yields a very interesting result: there is a parametrically light state,
the existence of which emerges only when $\r_{\ast}$ is large. The precise numerical values of the masses are slightly different
from those in~\cite{ENP} but qualitatively the same.
 This state does not exist for the classical
linear dilaton solution~\cite{MN}, and indeed our results agree  (within numerical accuracy) with those of~\cite{BHM}
in the $\r_{\ast}\rightarrow 0$ limit.
All of this means that we have obtained results which are in qualitative agreement with~\cite{ENP}.

\section{Walking backgrounds with constant-dilaton asymptotics}

The walking backgrounds of class 2 are very different.
In the far-UV the exponential behavior of $P\sim e^{4\r/3}$
is the consequence of the fact that in the dual field theory
there is a dimension-8 operator.
The subtle relation between this fact and the baryonic VEV, and the fact that it is possible to UV-complete the dual theory,
by algebraically constructing new solutions of type IIB in which $F_5$ and $B_2$ are non-trivial,
is explained in~\cite{stringWTCbaryonic}.
We will not further pursue this line, which requires a dedicated study.

\begin{figure}[t]
\begin{center}
\begin{picture}(500,140)
\put(30,10){\includegraphics[height=4.5cm]{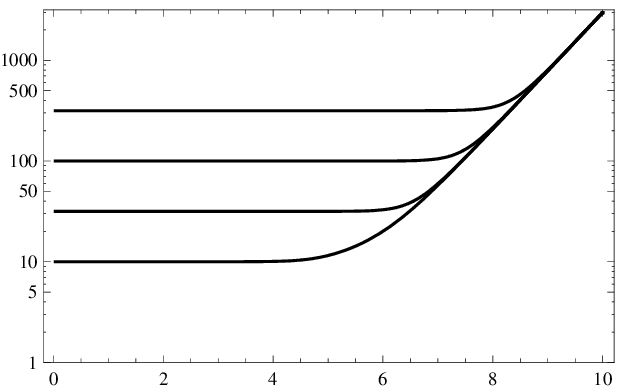}}
\put(270,10){\includegraphics[height=4.5cm]{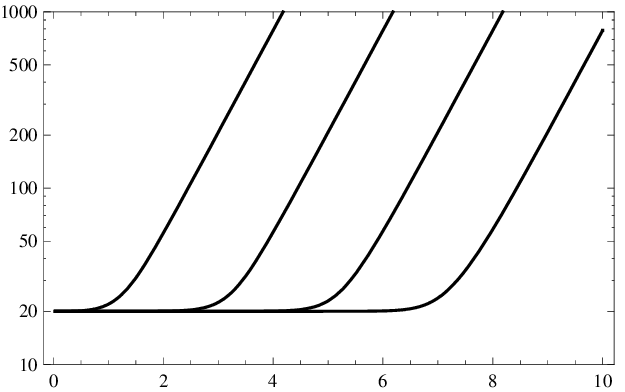}}
\put(10,120){$P$}
\put(255,120){$P$}
\put(210,0){$\rho$}
\put(450,0){$\r$}
\end{picture}
\caption{The function $P$, solution to the master equation, for various examples of solutions in class 2 (notice the logarithmic scale).
In the left panel a subclass having all the same leading asymptotic behavior in the UV, but different values of $P_0=P(0)$.
In the right panel, a set of solutions for $P$ having the same value of $P_0$, but different UV-asymptotics.}
\label{Fig:exponentials}
\end{center}
\end{figure}
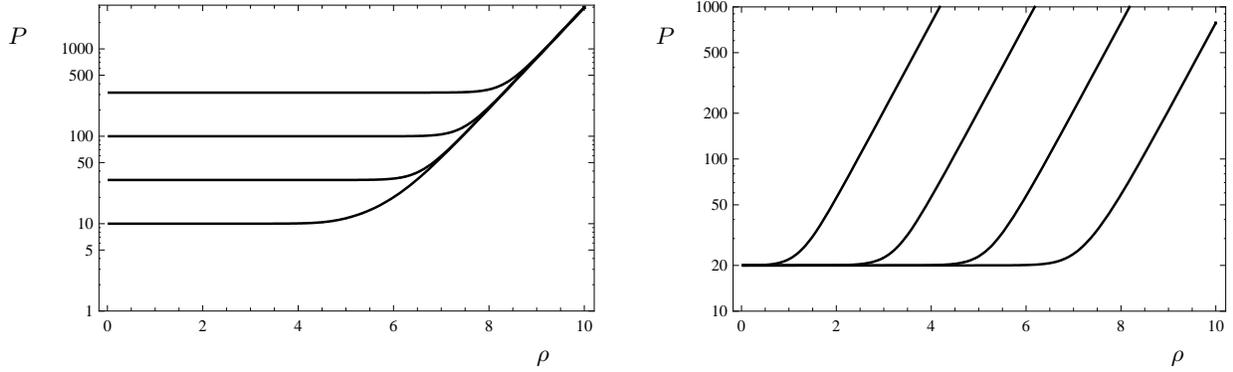

Our attention  will focus on the UV-cutoff dependence of the glueball 
spectrum, and on the effect of the rigorous boundary conditions. 
To do so, we start from the UV-expansion of the relevant quantities appearing in the bulk equations for the fluctuations.
In the UV, these solutions behave as
\SP{
	P = 3c_+ e^{4\rho/3} + \frac{4}{3c_+} \left(\rho^2 - \rho + \frac{13}{16} \right) e^{-4\rho/3} - \left(8c_+ \rho + \frac{c_-}{192c_+^2}\right) e^{-8\rho/3} + \mathcal O\left(e^{-4\rho}\right).
}
The two integration constants $c_+$ and $c_-$ characterize completely the background.
The former is related, on the field theory side of the  duality,
 to the  insertion of a dimension-8 operator, the latter to the appearance of a dimension-6 VEV which dominates the IR
physics (which is ultimately the reason why one might expect a light dilaton in the spectrum). For illustration purposes we show in Fig.~\ref{Fig:exponentials} a sample of possible solutions $P$ obtained by varying $c_+$ and $c_-$,
or equivalently $P_0$ and $\r_{\ast}$.
 
With these expansions, we find that
\beq
	S^a_{\ b} = \left(
\begin{array}{cccccc}
 \frac{8}{3} & 0 & 0 & 0 & 0 & 0 \\
 0 & \frac{8}{3} & 0 & 0 & 0 & 0 \\
 0 & 0 & \frac{8}{3} & 0 & 0 & 0 \\
 0 & 0 & 0 & \frac{8}{3} & 0 & 0 \\
 0 & 0 & 0 & 0 & \frac{8}{3} & 0 \\
 0 & 0 & 0 & 0 & 0 & 0
\end{array}
\right) + \mathcal O\left( e^{-4\rho/3} \right), \ \ 
	T^a_{\ b} = \left(
\begin{array}{cccccc}
 \frac{16}{9} & 0 & 0 & 0 & 0 & 0 \\
 0 & -\frac{16}{5} & -\frac{16}{15} & 0 & 0 & 0 \\
 0 & -\frac{32}{5} & -\frac{32}{15} & 0 & 0 & 0 \\
 0 & 0 & 0 & 0 & 0 & 0 \\
 0 & 0 & 0 & 0 & \frac{4}{3} & 0 \\
 0 & 0 & 0 & 0 & 8 & -4
\end{array}
\right) + \mathcal O\left( e^{-4\rho/3} \right),
\eeq
in which all the non-zero entries are ${\cal O}(1)$. However, the factor multiplying $q^2$ in Eq.~\eqref{eq:flucdiffrho} behaves as
\SP{
	e^{2k} = 2c_+ e^{4\rho/3} + \mathcal O\left( e^{-4\rho/3} \right),
}
growing without bound in the UV. This means that for any value of $q^2$, the fluctuations become oscillatory in the far UV, so that the spectrum only consists of a continuum, and therefore, contrary to what happens in class 1, there is no range of $q^2$ in which a discrete spectrum can be obtained.
The only way to get a discrete spectrum of bound states is by assuming that a UV-cutoff is physical.
But then the results of the calculations will depend explicitly on this cutoff, and hence they cannot be interpreted as the spectrum
of a UV-complete strongly-coupled confining field theory.
As anticipated, a possible way to provide a UV-completion exists, and the study of the spectrum
in this case will be presented elsewhere.

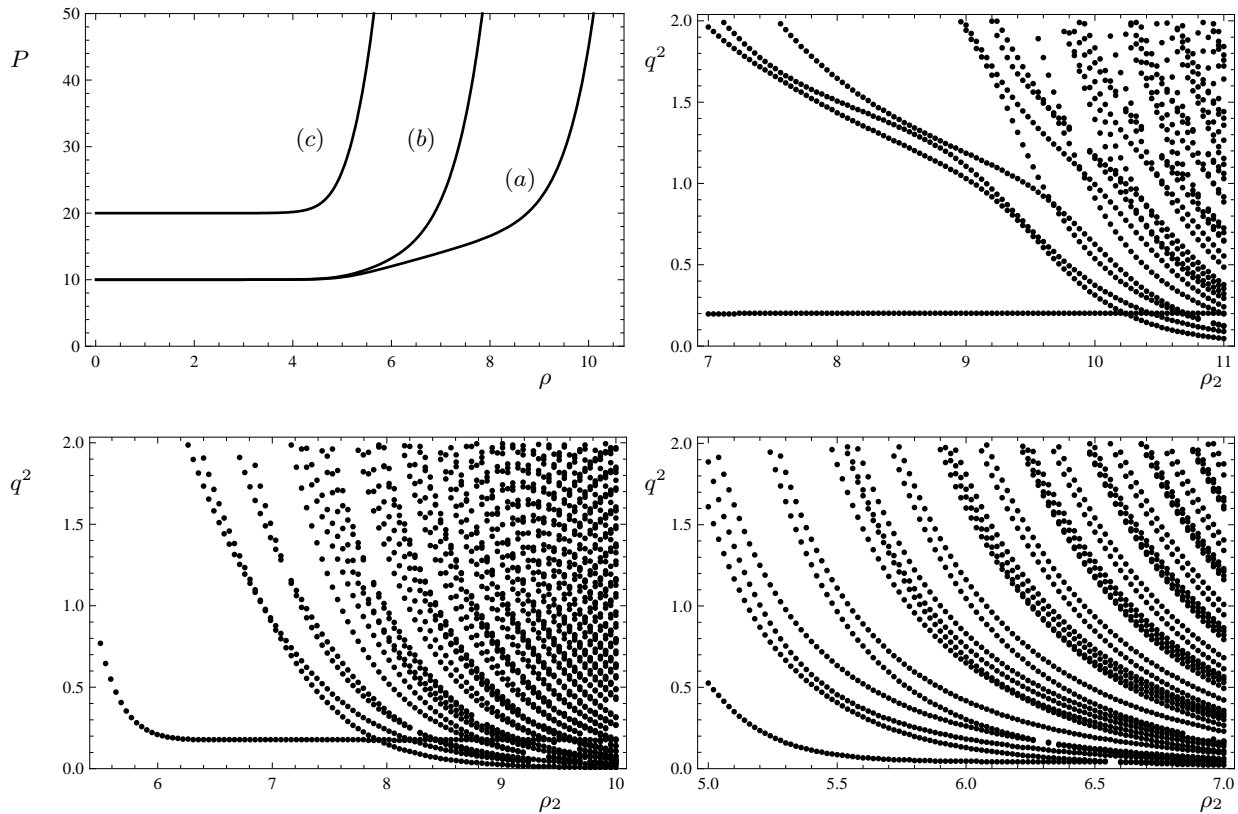
\begin{figure}[t]
\begin{center}
\begin{picture}(500,300)
\put(30,170){\includegraphics[height=4.9cm]{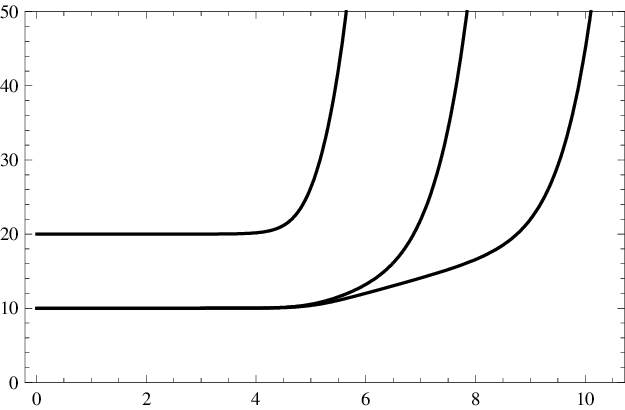}}
\put(260,10){\includegraphics[height=4.8cm]{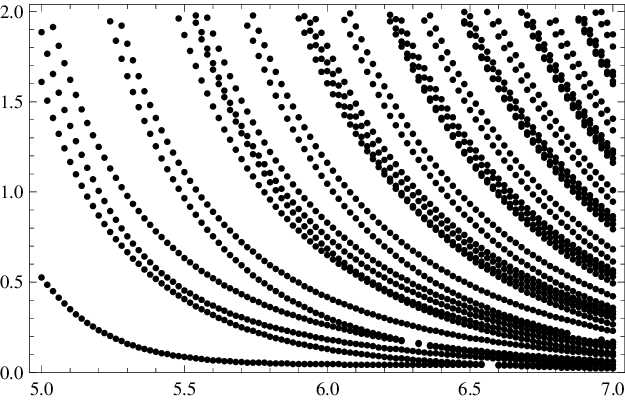}}
\put(30,10){\includegraphics[height=4.8cm]{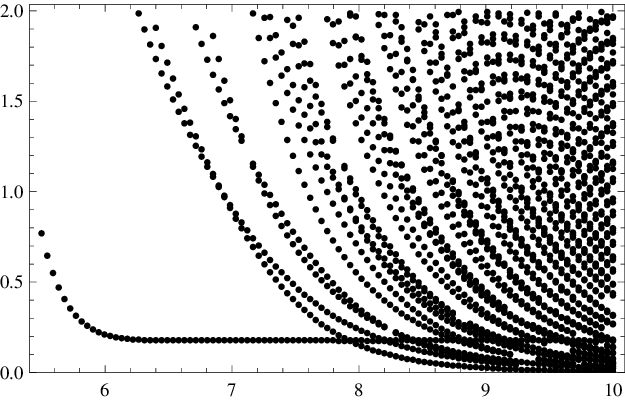}}
\put(260,170){\includegraphics[height=4.8cm]{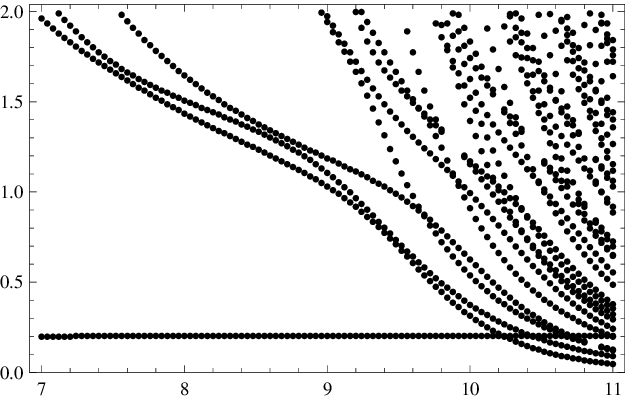}}
\put(250,125){$q^2$}
\put(10,125){$q^2$}
\put(210,5){$\r_2$}
\put(460,5){$\r_2$}
\put(250,285){$q^2$}
\put(10,285){$P$}
\put(210,165){$\r$}
\put(460,165){$\r_2$}
\put(197,240){$(a)$}
\put(160,255){$(b)$}
\put(118,255){$(c)$}
\end{picture} 
\caption{Spectrum obtained for examples of $P$ in class 2.
The upper-left panel shows the three choices of $P$ we analyze, labelled as $(a)$, $(b)$ and $(c)$. 
The other three panels show the spectrum of $q^2$ obtained for these solutions, as a function of $\r_2$.
Example $(a)$ is shown in the upper-right panel, example $(b)$ in the lower-left and 
example $(c)$ in the lower-right panel.
notice that all the spectra degenerate to a continuum for large $\r_2$,
and that in some cases an anomalously light state is present as long as $\r_2$ is kept fixed and small.
}
\label{Fig:exponentials2}
\end{center}
\end{figure}

For completeness, we computed the spectrum for three particular solutions in this class, by varying the value used for the cutoff.
We report the results in Fig.~\ref{Fig:exponentials2}. Notice how the whole spectrum depends strongly on $\r_2$ when $\r_2$ is very large,
to show explicitly that the discrete spectrum of bound states actually does not exists, but is an artifact of the UV-regulator.
But notice also that if one interprets $\r_2$ as a physical scale, and keeps it fixed (and smallish), 
the result is a spectrum in which one state is anomalously light,
at least is some region of the parameter space.
Notice also that the numerical value of the light mass obtained in this way, for the examples $(a)$ and $(b)$ in Fig.~\ref{Fig:exponentials2}
agrees reasonably well with the mass computed for the same value of $P_0$ in Fig.~\ref{Fig:spectrum1},
but then this result looses meaning when taking $\r_2$ large, because the state is overwelmed by the 
fact that the heavy states all become light.
In the case of example $(c)$, the situation is more subtle: for small values of $\r_2$ is it not possible to really gauge whether
there is a sense in which a light state is present. 
One important difference with~\cite{ASW} is in the fact that we used the rigorously derived boundary conditions from~\cite{EP}.
However, the existence or not of a light state in all of these backgrounds cannot be established here, 
but requires redoing the analysis for backgrounds in which the UV-asymptotic behavior is more tame~\cite{stringWTCbaryonic}.

There is also another significant difference between our analysis and the one in~\cite{ASW}.
To understand this, we need to digress, and explain to the reader more in detail the meaning
of the integration constants in $P$, in relation to the properties of the master equation.
By looking at the master equation, and the relation between $P$ and the functions appearing in the supergravity ansatz,
one finds that obtaining a smooth solution requires imposing the requirement that  $P$ and $P^{\prime}$ be monotonically increasing,
and satisfy always $P>Q$.
By inspection, it turns out that actually one has to require $P\geq 2N_c \r$.
Also by inspection, one finds that there are only three possible behaviors admissible for $P$: it can be 
approximately constant, linear with $\r$, or exponential with $P \sim e^{4\r/3}$. 
Together with the requirements of monotonicity, this means that any solution can be approximated 
by three sections over which $P(\r)$ is approximately constant at small-$\r$, linear at intermediate-$\r$,
and exponential at large-$\r$. 
Such is the behavior of  example $(a)$ studied and shown in Fig.~\ref{Fig:exponentials2}.
One or more of these behaviors may be absent, provided
one always has $P > 2N_c \r$, such as for $(b)$ and $(c)$ in Fig.~\ref{Fig:exponentials2}.
The walking region is the one in which $P$ is approximately constant, as explained in~\cite{NPP}.

In~\cite{NPP} a special class of solution was identified, in which the linear behavior is absent. The general solution for 
$P$ was shown to be written as a power expansion in a parameter $c$ such that 
\beqs
P&=&c P_1 +\frac{1}{c} P_{-1} + \frac{1}{c^3} P_{-3} +\cdots\,,
\eeqs
and in Figs.~1, 2 and 3 of~\cite{NPP} the goodness of such an approximation was gauged by comparing 
 to the exact numerical solution of the master
equation. The result is that for any physical quantity that does not depend on the 10-dimensional dilaton $\phi$ the agreement
between the leading order approximation $P\sim c P_1$ and the exact result is  good,
but the convergence of the series is slow for the dilaton, in which the leading-order approximation is 
visibly very different from the exact result, even at the qualitative level ($\phi$ appears to be non-monotonic in the approximate expression). 
Hence, any calculation where one needs to use the dilaton requires using the numerical result,
unless one dials to unreasonably large values the constant $c$, such that for all practical purposes the dilaton becomes exactly constant.
In particular, the calculation of the spectrum of scalar excitations is one such delicate case, because the dilaton appears 
in a very non-trivial way both in the sigma-model metric and in the potential.

The physical meaning of the parameter $c$ in this expansion is closely related to the dimension-2 baryonic VEV.
A complete discussion of this subtle issue can be found for example in~\cite{stringWTCbaryonic} (and relies on results
from~\cite{quivers,dimensions}). Large values of $c$ 
are related (again, non-trivially) to taking small values of this VEV.
At a more technical level, one finds that for a given $P_1(\rho)$ there exists a minimum value of $c$ such that the constraint $P\gsim 2N_c \r$ 
is satisfied for all $\r$~\cite{NPP}: in this case, $P$ becomes very close to the linear dilaton solution, at least 
for some intermediate region of $\r$. The case of the solutions of class 1 is such that this minimum value is realized, in such a way that 
asymptotically $P \simeq 2 N_c \r$ in the UV.
On the contrary, large values of $c$ parametrically suppress the dual gauge coupling 
defined according to~\cite{gauge}, and are realized by solutions that are always very 
different from the solutions in~\cite{MN} and (more relevant to this paper) those in~\cite{ENP}.

All of this means that the regime in which the approximation $P\simeq c P_1$ can be used is the opposite region of parameter space
in respect to the one in which the analysis in~\cite{ENP} identified a light composite scalar, and hence the analysis in~\cite{ASW},
which adopts this approximation throughout the whole paper, cannot be compared to~\cite{ENP}.

\section{Conclusions}

In this paper we computed the spectrum of glueballs of the theories the gauge-gravity
dual of which is represented by  walking solutions to the wrapped-$D5$
system in Type-IIB. We considered two possible classes of solutions, which we name class 1 and class 2,
and that differ by the UV-asymptotic behavior.
We applied in full rigor the tools of consistent truncation and of the gauge invariant formalism
developed in~\cite{BHM}, supplemented by the
 boundary conditions of~\cite{EP}.
 
 In class 1 we find qualitative agreement with the literature: in spite of the fact that the boundary conditions
 in~\cite{ENP} had been chosen on the basis of regularity and superficially look very different from those constructed rigorously
 in~\cite{EP} and applied here, both analyses show that there is a parametrically light scalar state, the mass of which
 is suppressed by the length of the walking region. It is still to be proven that this is a dilaton,
 but because it is known that the walking region is related to the presence of a  VEV for a 
 dimension-6 operator in the dual theory, it is possible that this is actually the case.
 This question has to be investigated further.
 
For class 2, we perform the analysis of the spectrum for a sample of possible solutions.
 First of all, because $P$ diverges exponentially in the UV, and hence a higher-dimensional operator is present in the dual
 theory, spoiling its UV-completeness, it is not possible to remove the UV-cutoff from the calculation,
 and hence all the results depend explicitly on such a cutoff, making their interpretation questionable.
Secondly, even if one takes the UV cutoff to be fixed, a light state is present in
the spectrum in the interesting range of parameter space where the comparison  to backgrounds of class 1 can be made~\cite{ENP}.

\vspace{1.0cm}
\begin{acknowledgments}
MP would like to thank the National Chiao-Tung University, Hsinchu, Taiwan and the 
National Taiwan University, Taipei, Taiwan  for hospitality during the completion of this work.
The work of MP is supported in part by WIMCS and by the STFC grant ST/J000043/1.
\end{acknowledgments}

\appendix
\section{Goldberger-Wise mechanism and boundary conditions}

We consider here a simple exercise, as a commentary on the use of the boundary conditions in Eq.~(\ref{eq:BCs2}).
We reconsider one of the examples in~\cite{EP}, namely the spectrum of the simplest possible realization
of the Goldberger-Wise mechanism, already discussed and well-known in the literature~\cite{dilaton5D}.
We do so in order to illustrate how dangerous it is to use naive choices of boundary conditions,
in place of the rigorously inferred Eq.~(\ref{eq:BCs2}).

The model is defined by a sigma-model with only one scalar $\Phi$, having canonical kinetic term,
and superpotential
\beqs
W(\Phi)&=&-\frac{3}{2}-\frac{\Delta}{2}\Phi^2\,,
\eeqs
with the potential given by 
\beqs
V&=&\frac{1}{2}\left(\frac{\partial W}{\partial \Phi}\right)^2 - \frac{3}{4} W^2\,.
\eeqs
The background solution we are interested in is
\beqs
\Phi&=&\Phi_1 e^{-\Delta r}\,,\\
A&=&r-\frac{1}{6}\Phi_1^2e^{-2\Delta r}\,,
\eeqs
and for simplicity we set $r_1=0$ and keep it fixed.
We will also keep $r_2\gg r_1$ fixed.

Following~\cite{dilaton5D,EP}, we take $\Phi_1\in {\cal O}(1)$, and assume that $r_2\gg 1$, and then
solve the bulk equations for the fluctuations in the limit in which $q^2=0$.
This yields 
\beqs
\mathfrak a(r)&=&c_1 e^{-\Delta r} +c_2 e^{-(4-\Delta)r}\,,
\eeqs
with $c_1$ and $c_2$ arbitrary real integration constants.
We then plug the result into the IR and UV boundary conditions
and solve for $c_1/c_2$ and $q^2$. The result is well known~\cite{dilaton5D}:
\beqs
q^2&\simeq&\frac{4\Delta^2(\Delta-2)\Phi_1^2}{3(-1+e^{2(\Delta-2) r_2})}\,.
\eeqs
Notice that in~\cite{EP} this result was also tested numerically, without making use of any of the approximations 
mentioned above, obtaining excellent agreement for choices of parameters for which the approximations
yielding this result hold.

Some comment about the physical meaning of this result will be useful later.
When $\Delta>2$, the background scalar represents a VEV in the dual,
otherwise conformal, field theory. The presence of a UV cutoff is the only 
explicit source of breaking of scale invariance, and hence in this case the mass of this lightest state
is parametrically small.
In the case $|\Delta|\ll 1$, the scalar represents the insertion in the dual theory of a quasi-marginal operator.
The explicit breaking of scale invariance is hence small, 
while the presence of the IR cutoff induces spontaneous breaking. Again, one finds a parametrically light scalar.
The last case, when $\Delta<2$ is not small represents the generic case in which the effects of 
explicit symmetry breaking are larger than the  effects of spontaneous breaking, and the resulting mass is not small.
This is also the case in which the approximations we made break down (unless $\Phi_1 \ll 1$), as shown in~\cite{EP} for example.
All of this is in perfect agreement with field-theory expectations.

Let us look more in detail at the practical effects of the UV boundary condition.
We can write it explicitly (at leading order in the approximations we are making):
\beqs
\left.\frac{}{}-\frac{2}{3}\Delta^2\Phi_1^2\frac{e^{(2-2\Delta)r}}{q^2}\left(\partial +\Delta\right) \mathfrak a \right|_{r_2}&=&\mathfrak a \left.\frac{}{}\right|_{r_2}\,.
\eeqs
Noticing that $(\partial_r+\Delta) \mathfrak a=(-4+2\Delta)c_2e^{-(4 - \Delta)r}$, one sees that in the limit in which $r_2$ becomes parametrically large,
this boundary condition reduces to the requirement that the subdominant asymptotic solution is kept, while setting the dominant solution to zero.
This would be the result also if we were to impose the simpler $\mathfrak a(r_2)=0$ boundary conditions.
The advantage of using   Eq.~(\ref{eq:BCs2}) is not clear in the case of one scalar and of a background which is known in closed form,
but when many scalars are present, they mix, and the background is known only numerically, this is clearly a much more effective way of writing the boundary conditions. In particular, in the case of mixing between many scalars, it may be difficult to give a precise meaning to 
what it means to be dominant or subdominant for any of the possible asymptotic solutions, since in each scalar there will be more than two
independent coefficients appearing at various orders (because of mixing, again).
Also, this equivalence of choices is true only provided the $r_2\rightarrow +\infty$ limit can be taken.
If one wants to interpret $r_2$ as a physical cutoff (kept fixed) the three choices yield different results, which converge only for large $r_2$.
We remind the reader that Eq.~(\ref{eq:BCs2}) descends directly from the requirement that the variational principle be well-defined for the
full five-dimensional action, and hence this is the correct choice (up to the ambiguity in the coefficients $m_i^2$ discussed in~\cite{EP},
and not relevant here).

Let us now perform the exercise of computing again the mass of the lightest scalar state, but this time we will make use of an 
arbitrarily modified form of the IR boundary conditions. For simplicity, and considering that we have only one scalar
for which a semi-analytical solution is known, we impose $\mathfrak{a}(r_2)=0$, following the indications discussed 
earlier, and restrict our attention to very large values of $r_2$.
The form of the IR boundary conditions we want to test is controlled by two arbitrary parameters $\alpha$ and $\beta$, according to 
\beqs
\left.\frac{}{}-\frac{2}{3}\Delta^2\Phi_1^2\frac{e^{(2-2\Delta)r}}{{q}^2}\left(\alpha\partial +\beta\Delta\right) \mathfrak a \right|_{0}
&=& \beta\left.\frac{}{}\mathfrak a\right|_{0}\,.
\eeqs
With this choice, $\alpha=0$ means Dirichlet boundary conditions, while $\beta=0$ means Neumann boundary conditions.
After imposing the UV boundary conditions, and taking again $r_2\gg 1$, we find
\beqs
{q}^2&=&\frac{2\Delta^2\Phi_1^2}{3}\left(2\frac{\alpha}{\beta}-\Delta+\frac{\alpha}{\beta}(\Delta-2){\rm cotanh}\left(\frac{}{}(\Delta-2) r_2\right)\right)\,.
\eeqs
For generic values of $\alpha$ and $\beta$, we see that ${q}^2\in {\cal O}(1)$, hence failing to reproduce correctly the behavior
expected for $\Delta>2$. By making use of the expansion ${\rm cotanh}(x)\sim 1+2e^{-2x}$ we have (for $\Delta>2$)
\beqs
{q}^2 &\simeq& \frac{2(\alpha-\beta)\Delta^3\Phi_1^2}{3\beta}
\,+\,\frac{4\alpha}{3\beta}\Delta^2(\Delta-2)\Phi_1^2e^{-2(\Delta-2)r_2}\,.
\eeqs
Clearly, the first term is parametrically suppressed only for $|\Delta|\ll 1$, but not for $\Delta>2$.
The second term is parametrically suppressed when $\Delta>2$, but in order to obtain the light state
one must then cancel the first term, which requires $\alpha=\beta$,  which is the choice of boundary conditions we adopted in this paper!
Notice in particular that $\alpha=0$ corresponds to Dirichlet boundary conditions, and in this case
there cannot be light states unless $\Delta$ is very small. Even worse, taking $\beta=0$ (Neumann) yields to a singular behavior, which
just means that there is no light state for any possible choice of the parameters.

In conclusion, this simple little exercise shows that the boundary conditions we adopted in the main calculations of this and related papers, namely Eq.~(\ref{eq:BCs2}), reproduce the physical results (as we already knew from~\cite{dilaton5D,EP}). Imposing purely Neumann or Dirichlet boundary conditions 
yields unphysical results.
Ultimately, the reason for this is that the correct boundary conditions have been derived by imposing 
consistency with the variational problem.


\end{document}